\documentclass[a4paper]{jpconf}
\usepackage{graphicx,color}
\usepackage{lmodern}
\usepackage{amsmath}
\usepackage{siunitx}
\usepackage{bbm}
\usepackage[breaklinks=true,colorlinks=true,linkcolor=blue,urlcolor=blue,citecolor=blue]{hyperref} 

\newcommand{\lambdabar}{{\mkern0.75mu\mathchar '26\mkern -9.75mu\lambda}}

\begin{document}

\title{Does universality of free-fall apply to the motion of quantum probes?}

\author{Luigi Seveso}
\address{Quantum Technology Lab, Dipartimento di Fisica, Universit\`a degli Studi  di Milano, I-20133 Milano,  Italy}
\ead{luigi.seveso@unimi.it}

\author{Valerio Peri$^{1,2}$}
\address{$^1$Quantum Technology Lab, Dipartimento di Fisica, Universit\`a degli Studi  di Milano, I-20133 Milano,  Italy}
\address{$^2$Institute for Theoretical Physics, ETH Zurich, 8093 Z\"urich, Switzerland}
\ead{periv@student.ethz.ch}

\author{Matteo G. A. Paris$^{1,2}$}
\address{$^1$Quantum Technology Lab, Dipartimento di Fisica, Universit\`a degli Studi  di Milano, I-20133 Milano,  Italy}
\address{$^2$INFN, Sezione di Milano, I-20133 Milano, Italy}
\ead{matteo.paris@fisica.unimi.it}

\begin{abstract}
Can quantum-mechanical particles propagating on a fixed spacetime background be 
approximated as test bodies satisfying the weak equivalence principle? We ultimately answer 
the question in the negative but find that, when universality of free-fall is assessed locally, a nontrivial agreement between quantum mechanics and the weak equivalence principle exists. Implications for mass sensing by quantum probes are discussed in some details. 
\end{abstract}

\section{Introduction}
The weak equivalence principle (WEP) is arguably the most basic property of the classical gravitational field. It states that the trajectories \ {of a particle subject only to gravitational forces are completely determined} by the local spacetime geometry \cite{will1993theory,di2015nonequivalence}. Strictly speaking, the WEP applies only to test-bodies, i.e. particles which are sufficiently light so that their back-action on the external gravitational field is negligible, and sufficiently small so that the effects of tidal forces on them are inconsequential. For such probes, the WEP states that their trajectories under gravity are independent 
of the internal properties, \ {most remarkably} their masses. 

In general relativity, the WEP is directly encoded into the specific form taken by the action functional for a 
free point-like material particle. The action $S(a,b)$ for the propagation of a particle of mass $m$ between 
the two events $a$ and $b$ is given by
\begin{equation}\label{action}
S(a, b) = -mc \int_a^b ds\;, 
\end{equation}
where $ds=\sqrt{g_{\mu\nu} dx^\mu dx^\nu}$ is the metric element. Since the mass appears as a 
multiplicative factor, it does not enter the equations of motions. \ {As a consequence, particles of 
different masses follow the same trajectories given the same initial conditions. This is the content of 
the WEP, as iconically represented by the Leaning Tower of Pisa experiment, supposedly performed 
by Galileo Galilei in 1589 \cite{galp1,galp2}, as well as its repetition on the moon by astronaut 
David Scott during the Apollo 15 mission \cite{apo15}. }

The classical gravitational theory thus undisputably predicts that
test-bodies move universally when in free-fall. However, it is a
logically different question whether the behavior of real physical
particles may, in fact, be approximated with that of test-bodies. The
question is especially relevant when the particle's propagation exhibits
quantum features
\cite{greenberger1968role,anandan1980hypotheses,sonego1995there,chryssomalakos2003geometrical,okon2011does},
i.e. if quantumness of the particle probing an external gravitational
field is taken into account.  \ {Is the WEP satisfied in the
quantum regime?} In this paper, we address the conceptual 
aspects of the problem and, in particular, we devote our analysis to
\begin{enumerate}
\item[1.] \textbf{discussion of the status of the WEP within quantum mechanics:} in Section \ref{qmWEP}, 
we review some standard objections against the possibility of extending the WEP of classical physics to quantum particles;
\item[2.]  \textbf{formulation of a WEP for quantum probes:} while with classical particles one can sensibly 
talk about trajectories, quantum particles are described by wavefunctions. Therefore, a suitable reformulation of the WEP that applies also to quantum probes is in order. \ {Our proposal, based on quantifying the 
information extractable from position measurements, is described in details} in Section \ref{wepquantum};
\item[3] \textbf{results for uniform and non-uniform gravitational fields:} in Section \ref{qWEPuni}, we 
discuss to what extent quantum probes obey the WEP according to our information-theoretic 
formulation, pointing out in particular the differences between uniform and non-uniform gravitational fields.  
\end{enumerate}

\ {Besides the foundational aspects, our analysis may be of interest 
to address experiments involving gravity and particles with 
an inherently quantum-mechanical behaviour, e.g. Bose-Einstein condensates
and neutron interferometry.}

\section{Quantum mechanics and the WEP} \label{qmWEP}
Let us here review the status of the WEP in classical and 
quantum physics, with a focus on their non-relativistic limits. 
The Newtonian theory of gravity implies that a test-body of 
mass $m$ in an external gravitational potential 
$\varphi(\textbf{x})$ obeys the equation of motion 
\begin{equation}\label{lapleq}
m\, \ddot{\textbf{x}} = - m\, \nabla \varphi(\textbf{x})\;.
\end{equation}
The mass simplifies and, as a result, the solutions of 
\eqref{lapleq} do not depend on $m$.  

\ {Moving to the quantum case,} one may incorporate gravity at the
semiclassical level by starting from the relativistic Klein-Gordon
equation for a massive spinless boson in a fixed metric background
\cite{kiefer1991quantum}. Then, upon taking the non-relativistic limit,
one recovers a  Schr\"odinger-like equation with the gravitational
potential energy $m\, \varphi(\textbf{x})$ of the probe contributing an
extra term to the Hamiltonian,  
\begin{equation}\label{scheq}
i\hbar \,\partial_t \psi(\textbf x,\,t) =  -\frac{\hbar^2}{2m}\,\Delta\psi(\textbf{x},\,t) + m\, \varphi(\textbf{x})\,\psi(\textbf{x},\,t)\;. 
\end{equation}
The particle's mass appears explicitly through the ratio $\hbar/m$. The
solutions of \eqref{scheq} thus depend parametrically on the same ratio.
It follows that quantumness of the state describing the probing particle
is directly linked to the appearance of mass-dependent effects in its
propagation. In other words, the dynamical evolution of the wavefunction
for a particle falling in a gravitational field contains non-trivial
information about its mass. This is because we may take repeated
measurements on identically prepared probes and, since the quantum state
of the system is mass-dependent, the statistics of the measurements
would also be, in general, mass-dependent.  
\ {
At a first sight, such a conclusion appears to be at odd with the
essence of the WEP, which dictates that no information about the mass of
a freely-falling particle is available to an observer monitoring its
trajectories.   As we will see, the WEP is indeed clashing with quantum
mechanics in the general case of a nonuniform field,  whereas a
nontrivial agreement between quantum mechanics and the WEP exists
for uniform fields, i.e. locally.}

\subsection{Quantifying information on a 
parameter: quantum parameter estimation theory}

\ {Information about the value of a certain parameter may be
extracted from the system under investigation by performing 
measurements. This intuitive concept may be made more 
precise in the framework of quantum parameter estimation theory.} 
In order for this contribution to be self-contained, we present a brief
overview of its fundamentals. A more extensive review can be found in
\cite{paris2009quantum}, together with the more in-depth references
\cite{helstrom1976quantum,holevo2003statistical,hayashi2006quantum}. 

Quantum parameter estimation theory provides the tools to quantify the
precision achievable in any estimation procedure aimed at inferring the
value of a certain physical parameter $\lambda$, taking into account the
limitations imposed by quantum mechanics. As a matter of fact, there are
many quantities in quantum systems which do not formally correspond to
proper physical observables (e.g. coupling constants). As such, they are
not directly measurable, even in principle. In these cases, one has to
resort to an indirect estimation strategy, i.e. one measures some
observable of the system, which is experimentally accessible, and whose
outcomes' statistics depends on the parameter, and then extracts an
estimate through a suitable processing of the data. Therefore, there are
two distinct steps involved in any estimation task: 1. the choice of the
measurement to be performed and 2. the choice of data processing, i.e. 
the choice of how to elaborate statistically the experimental data to 
extract an estimate of the parameter. 

The last step is usually fulfilled by a locally unbiased estimator $\hat
\lambda$, which is a function of the measurement outcomes such that its
expectation value equals the true value of the parameter. That is,
$\mathbbm{E}_\lambda(\hat \lambda) = \lambda$, where
$\mathbbm{E}_\lambda$ denotes the expectation value taken with respect
to the probability distribution $p_\lambda(x)$ of the experimental data
$x=\{x_1, x_2,\dots, x_N\}$. 

Typically, one further requires the estimator to minimize the
expectation value of a suitable loss function. One of the most natural
choices is to minimize the expectation value of the square of the
deviations of the estimator with respect to the true value of the
parameter, $\mathbbm{E}_\lambda[(\hat\lambda-\lambda)^2]$, which for
locally unbiased estimator coincides with the variance $\text{Var}(\hat
\lambda)$ of $\hat \lambda$. It is then a well-known result of classical
statistics (that goes under the name of Cr\'amer-Rao theorem
\cite{cramer2016mathematical}) that the variance of any locally unbiased
estimator after $N$ repeated measurements is bounded from below by the
inverse of the Fisher information $F_X(\lambda)$ (multiplied by $N$),
\begin{equation}
\text{Var}(\hat\lambda)\geq \frac{1}{N\,F_X(\lambda)}\;.
\end{equation}

The bound is saturable, at least in the asymptotic limit $N\to \infty$
of a large number of measurements, and the corresponding optimal
estimator is said to be efficient. The Fisher information thus
quantifies, for a fixed choice of measurement scheme, the maximum amount
of information which can be extracted on some unknown parameter.
Explicitly, it is defined as the expectation value of the logarithmic
derivative squared of the probability distribution of the data,
\begin{equation}
F_X(\lambda) = \mathbbm{E}[(\partial_\lambda \ln p_\lambda)^2]\;.
\end{equation}  

Regarding the first optimization step, i.e. the choice of the measurement, one is interested in achieving the best possible estimation precision, which corresponds to the maximum possible Fisher information. The Fisher information can in fact be maximized over the choice of the measurement scheme, which yields as an upper-bound the quantum Fisher information $J(\lambda)$, 
\begin{equation}
J(\lambda) = \underset{\{X\}}{\max}\; F_X(\lambda)\;,
\end{equation}
where the maximum is taken with respect to measurements of all possible observables $X$ \cite{braunstein1994statistical}. Closed-form expressions for $J(\lambda)$ are  available. The quantum Fisher information encodes the ultimate bound to the achievable precision for any quantum-limited measurement.

\subsection{A quantum Galilean experiment}
Returning to the case of a freely-falling quantum probe under gravity, the reasoning below \eqref{scheq} suggests that the Fisher information will in general be non-zero when measurements of position are employed to estimate its mass. The same could also be true for other kinds of measurements, e.g. a measurement of momentum. However, since the WEP concerns the trajectories of particles, it is the Fisher information for position measurements to be relevant. Thus, from now on the parameter to be estimated is the mass of a probe in a gravitational field, and the probability $p_m(\textbf{x})$ of measuring the particle at position $\textbf{x}$ is the modulus square of the wavefunction $\psi(\textbf{x},t)$ which is a solution of equation \eqref{scheq} for appropriate initial conditions. 

\ {In order to quantitatively illustrate some of the observations made above}, let us work out a simple example. More general examples of quantum measurement strategies aimed at mass estimation under gravity, and their corresponding sensitivities, can be found in \cite{seveso2016quantum}. Let us consider a quantum free-fall experiment \'a-la Galileo with a quantum probe in a uniform gravitational field $\textbf{g} = - g\,\hat{\textbf{i}}$ ($\hat{\textbf{i}}$ is the unit vector along the $x$ axis). The initial preparation of the probe is described by the Gaussian wavefunction
\begin{equation}
\psi(x,\, 0) = \frac{1}{(\pi\,\Delta x^2)^{1/4}}\, e^{-\frac{1}{2}\,\left(\frac{x-h}{\Delta x}\right)^2}\;,
\end{equation}
which in the classical limit describes a particle localized over the scale $\Delta x$ around $x=h$. Under time evolution (setting $\hbar=1$), we have
\begin{equation}\label{wft}
\psi(x,\,t) = \frac{1}{(\pi\,\Delta x^2)^{1/4}}\,e^{-img^2 t^3/6}\,e^{-imgxt}\,\frac{1}{\sqrt{1+it/m\,\Delta x^2 }}\, e^{-\frac{1}{2(1+it/m\,\Delta x^2 )}\left(\frac{x-h}{\Delta x}\right)^2}\;.
\end{equation}
One can then compute the Fisher information on $m$ which can be extracted from the system through measurements of position, and compare it with the ultimate quantum limit set by the quantum Fisher information,
\begin{equation}
F_X(m) =\frac{2}{m^2}\,\left[1+\left(\frac{m\,\Delta x^2}{t}\right)^2\right]^{-2}\;, \;\qquad J(m) =  \frac{t^2}{2 m^4 \Delta x^4} + 2 g^2 t^2 \Delta x^2 + \frac{2g^2t^4}{m^2\, \Delta x^2} \;.
\end{equation} 
The quantum Fisher information grows like $t^4$ for large interrogation times. In contrast, the Fisher information saturates to a constant value, which means that by monitoring the trajectory of a freely-falling quantum probe one cannot arbitrarily improve the precision.

This last remark is valid when the probe is subject only to gravitational forces. If this is not the case, a measurement of position can perform significantly better. Let us consider the following setup, which reproduces the basic physics of neutron interferometry experiments \cite{colella1975observation}. A beam of particles with de Broglie wavelength $\lambdabar$ is split in two and propagates along the two arms of an interferometry, with one arm being located higher by an amount $h$ in the Earth's gravitational field. The beam propagating in the upper arm acquires an extra phase due to gravity, $\varphi_g = -m^2 g h L \lambdabar/\hbar^2$, where $L$ is the length of the horizontal arm of the interferometer. Then the two beams are recombined on a beamsplitter and a coarse-grained position measurement is made, i.e. it is measured whether the particle leaves the interferometer through one port or the other. The Fisher information on the particle's mass that can be extracted in this way is equal to $4\varphi_g^2/m^2$. It scales quadratically with $L$ and therefore likewise with the interrogation time of the experiment. Thus, if the particle probing the field is not in pure free-fall, monitoring of its trajectory allows to estimate its mass to any desired level of accuracy by increasing the interrogation time.\footnote{However, a single measurement would not suffice. It is still necessary that the Cr\'amer-Rao inequality be saturated, which in general requires a number of repetitions sufficiently large to be close to the asymptotic regime.} 

\begin{figure}[h]
\centering
\includegraphics[width=0.6\textwidth]{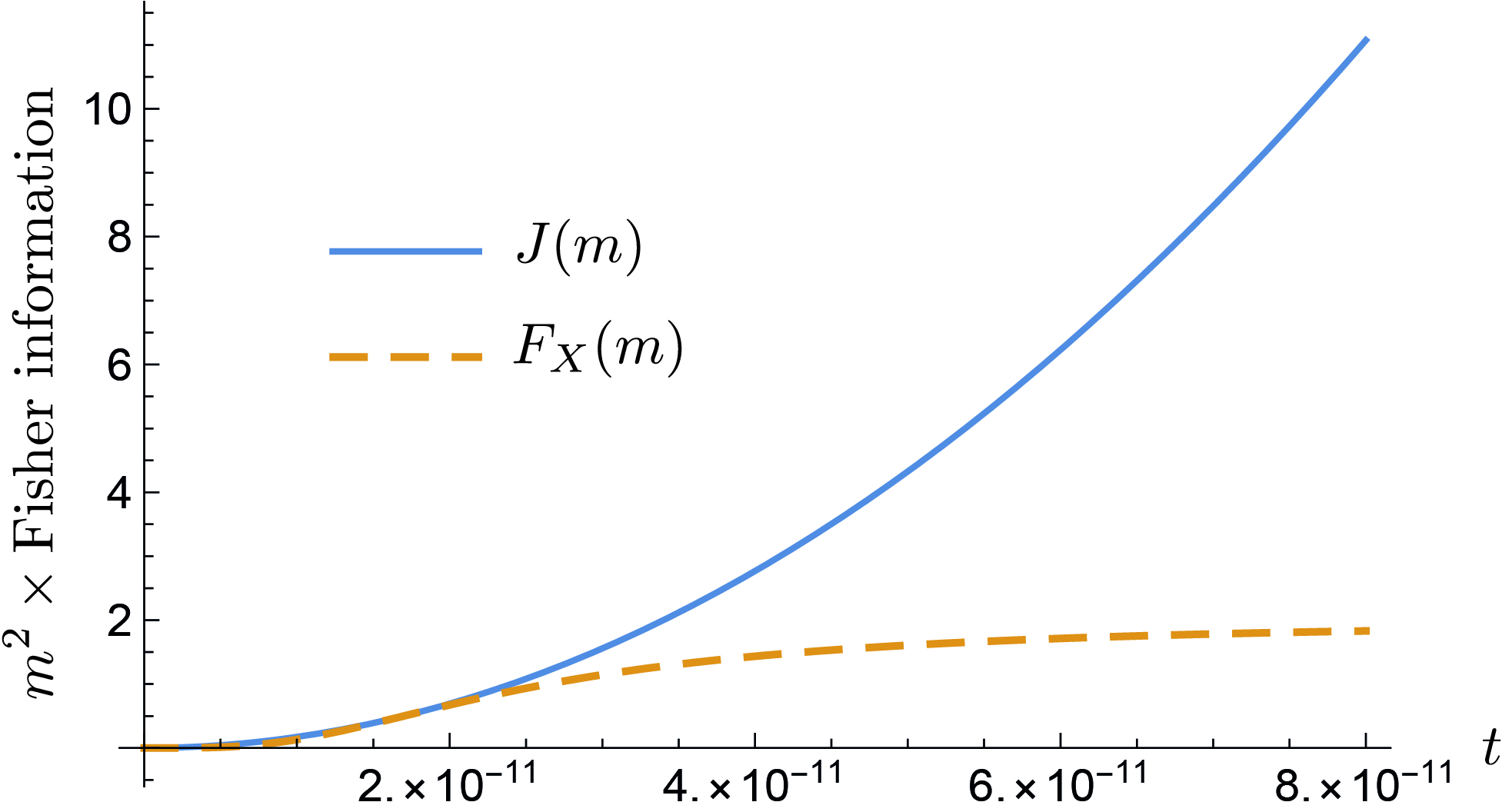}%
\caption{\label{label}\ {Comparison between the quantum Fisher information $J(m)$ (\emph{solid}) and the position Fisher information $F_X(m)$ (\emph{dashed}) for the mass of a freely-falling Gaussian probe  
in a uniform field. Both quantities are multiplied by $m^2$ to make them adimensional. For the purposes 
of this figure, the mass $m$ is taken to be the neutron's mass $m\simeq \SI{1.7e-27}{\kilogram}$ and 
the time $t$ is reported in seconds.}}
\end{figure}

\section{Information theory-inspired WEP for quantum probes?} \label{wepquantum}
The trajectories of classical test-bodies in free-fall carry no information about their masses. Vice versa, measurements of position on quantum probes allow to extract nonvanishing information. In particular, even in the point-like limit when $\Delta x\to 0$, $F_X(m)$ does not go to zero. The fact that the Fisher information does not vanish may be seen as evidence that a quantum probe cannot be approximated by a test-body. 

However, we are going to argue that such a conclusion is premature. In fact, the solutions of the Schr\"odinger equation depend on the ratio $\hbar/m$ also in the free case. Therefore, the propagation of a wavepacket is subject to mass-dependent dispersive effects even in the absence of any gravitational field. As a consequence, one would generically obtain a nonzero Fisher information by monitoring the probe's trajectory. 

In order to assess the validity of the test-body approximation for quantum probes, one should focus only on the information which is explicitly due to the gravitational coupling. In other words, one may split the Fisher information as
\begin{equation}
F_X(m) = F_X(m)|_{free} + F_X(m)|_{grav}\;,
\end{equation}
where $F_X(m)|_{free}$ is the Fisher information one would have if no gravitational field is present (for the same initial preparation of the probe), and $F_X(m)|_{grav}$ is the Fisher information which is  introduced by the external field. We will therefore say that quantum probes satisfy the WEP if the introduction of a gravitational field does not allow to extract any information $F_X(m)|_{grav}$, in addition to the amount $F_X(m)|_{free}$ which would already be present with no gravity. Vice versa, if it turns out that $F_X(m)|_{grav}$ is in general nonzero, then we will conclude that quantum probes cannot behave even in principle as test-particles. 

Such formulation of the WEP is sufficiently general to apply to properly quantum particles. It has an operative flavour (as it is based on the results of physical measurements), and is inspired by information theory concepts. In the next section, we discuss to what degrees it is actually satisfied by quantum mechanics.     

\section{Results for uniform and non-uniform gravitational fields} \label{qWEPuni}

We tackle first the case of a uniform gravitational field. Let $\psi(\textbf{x}, 0)$ denote the initial preparation of the probe at time $t=0$ and let $\textbf{g}$ be a uniform gravitational field corresponding to a potential $\varphi(\textbf{x})$. The state of the probe is then evolved according to the Hamiltonian $H = \textbf{P}^2/2m + m \varphi(\textbf{x}) $. The resulting state at general time $t$ is
\begin{equation}\label{state}
\psi(\textbf{x}, 0) = e^{-img^2 t^3/6}\, e^{-imt\varphi(\textbf{x})}\,\psi_{free}(\textbf{x}-\textbf{g}t^2/2,\, t)\;,
\end{equation}
where $\psi_{free}(\textbf{x},\,t)$ is the time-evolved wavefunction in the absence of any gravitational field and given the same initial condition. It is then a simple matter \cite{seveso2016can} to check that the Fisher information for the state \eqref{state} satisfies  $ F_X(m)=F_X(m)|_{free}$, i.e. $F_X(m)|_{grav}=0$. Thus, at least locally, no clash between quantum mechanics and the WEP is present. 

More interesting is the case of a non-uniform gravitational field. In this case it is not possible in general to obtain exact solutions of the Schr\"odinger equation \eqref{scheq}; however one can obtain approximate solutions by solving \eqref{scheq} through a BCH expansion \cite{serre2009lie} of the time-evolution operator, neglecting terms which are second order in the gravity gradient $\nabla \textbf{g}$ and its derivatives \cite{seveso2016can}. Then, one finds that the probability distribution of position measurements is modified in a mass-dependent way by the introduction of an external gravitational field. More precisely, 
\begin{equation}
|\psi(\textbf{x},\,t)|^2 \sim |\psi_{free}(\textbf{x}-\textbf{g}t^2/2 + \textbf{d}_m,\,t)|^2\;,
\end{equation}
where $\textbf{d}_m$ is a displacement which violates universality, i.e. it depends explicitly on the mass $m$ of the probing particle,\footnote{A few words on notation. For example, the writing $\textbf{x}\cdot \nabla \textbf{g}(\textbf{x})$ stands for the three-dimensional vector whose $j^{\text{th}}$ component is $x_i\partial_i g_j$, with repeated indices summmed over. There is no ambiguity over contractions of indices since $x_i\partial_i g_j=x_i\partial_j g_i$. The same considerations apply for other analogous expressions in the text.} 
\begin{equation}
\textbf{d}_m= \frac{t^2}{2}[\textbf{x}\cdot \nabla \textbf{g}(\textbf{x})-\textbf{g}(\textbf{x})]+\frac{t^3}{3m}\textbf{P}\cdot\nabla \textbf{g}(\textbf{x})+\frac{5 t^4}{48}\nabla \textbf{g}^2(\textbf{x})\;.
\end{equation} 
Notice that the mass-dependent terms vanishes identically for a uniform field. More generally, we expect our formulation of the WEP to be satisfied when such a term can be neglected compared to the term $-\textbf{g}t^2/2$, which would imply
\begin{equation}\label{cond}
\left|t^3\,\textbf{p}\cdot \nabla \textbf{g}/m\right|\ll |\textbf{g}t^2|\;.
\end{equation}
By substituting $\textbf{p}$ with its classical expression $m\textbf{g}t$ and approximating the gravity gradient as $\sim\textbf{g}/\ell$, where $\ell$ is the characteristic scale of variation of the gravitational field, one finds that \eqref{cond} is equivalent to the condition
\begin{equation}
|\textbf{g}t^2| \ll \ell\;,
\end{equation}
i.e. as long as the probe does not have the time to explore regions of the order of the curvature scale the WEP appproximately holds. In general, however, gravity gradients can encode non-trivial information about the mass of a particle into its wavefunction, so that $F_X(m)|_{grav}\neq 0$.   

\section{Conclusions}
\ {In conclusion, we have discussed the use of quantum parameter estimation theory to analyze the interplay between quantum mechanics and the WEP. We have shown that, besides being a fundamental tool for
the design of sensing experiments, the Fisher information may also provide a way to discuss whether quantum probes can be approximated as test-bodies. We found that, in general, this is not the case: the evolution of the wavefunction of a quantum probe is sensitive in a mass-dependent way to gravity gradients. This, in turn, means that the introduction of a gravitational field may improve precision in estimating the mass of a 
freely-falling particle by monitoring its trajectory. Nonetheless, locally, i.e. in the limit of a uniform field, our formulation of the WEP does hold.}

Let us emphasize that for classical probes, universality of free-fall holds in a stronger sense: a probe, devoid of any internal structure as considered here, but subject only to classical physics, would satisfy our formulation of the WEP irrespective of whether the external field is uniform or not. On the other hand, if the classical probe has an internal structure, e.g. it is spatially extended, then it is known \cite{geroch1975motion} that it does not in general move geodesically. \ {Our results show that a quantum probe, even if in principle devoid of any internal structure, behaves instead in analogy with a non-pointlike classical probe, because of the extended character of its wavefunction.} 

\ {
\ack This work has been supported by EU through the Collaborative 
Project QuProCS (Grant Agreement 641277) and by UniMI through the H2020 Transition Grant 15-6-3008000-625. 
}

\section*{References}
\bibliographystyle{iopart-num}
\bibliography{DICErefs}

\end{document}